\documentclass{mem}
\usepackage{natbib}\usepackage{txfonts}\usepackage{balance}
\usepackage{graphicx}
\usepackage[a4paper]{hyperref}
\idline{0}{0}
\begin{document}
\def\teff{$T\rm_{eff }$}
\def\kms{$\mathrm {km s}^{-1}$}

\title{
Binarity and multiperiodicity in high-amplitude delta Scuti stars
}

   \subtitle{}

\author{
A. \,Derekas\inst{1,2},
L. L. \,Kiss\inst{2},
B. \,Cs\'ak\inst{3},
J. \,Griffin\inst{2},
C. \,Lindstr\o m\inst{2},
Sz. \,M\'esz\'aros\inst{4},
P. \,Sz\'ekely\inst{5},
M. C. B. \,Ashley\inst{1}
and T. R. \,Bedding\inst{2}
          }

  \offprints{A. Derekas}

\institute{
School of Physics, Department of Astrophysics and Optics, University of New 
South Wales, Sydney, NSW 2052, Australia;
\email{derekas@physics.usyd.edu.au}
\and
School of Physics, University of Sydney, NSW 2006, Australia
\and
Harvard-Smithsonian Center for Astrophysics (CfA), 60 Garden Street, Cambridge, 
MA 02138, USA
\and
Department of Optics and Quantum Electronics, University of Szeged, Hungary
\and
Department of Experimental Physics and Astronomical Observatory, University of 
Szeged, Hungary
}

\authorrunning{Derekas}

\titlerunning{binary and multiperiodic HADS's}

\abstract{

We present our first results for a sample of southern high-amplitude $\delta$
Scuti stars (HADS), based on a spectrophotometric survey started in 2003. For
CY Aqr and AD CMi, we found very stable light and radial velocity (RV) curves;
we confirmed the double-mode nature of ZZ Mic, BQ Ind and RY Lep. Finally, we
detected $\gamma$-velocity changes in RS Gru and RY Lep.

}
\maketitle{}

\section{Introduction}

Both binarity and multiple periodicity offer independent methods to determine
fundamental physical parameters of oscillating stars. Our aim in this project
is to detect binarity or multiperiodicity (or both) in selected bright southern
HADS variables in order to combine binary star astrophysics with
asteroseismology. Currently, only a few HADS are known in binary systems (e.g.
Rodr\'iguez \& Breger 2001 listed six such stars), but there are more with
suspected binarity.

Here we present our first results for six southern stars. Previous analyses of
northern variables can be found in Kiss et al. (2002) and Derekas et al. (2003).

\section{Observations}

We carried out photoelectric, CCD photometric and medium-resolution
spectroscopic observations with four instruments at Siding Spring Observatory
(Australia) on approximately 50 nights between 2003 October and 2005 May. The
24$^{\prime\prime}$ telescope was used for BVI photometry; CCD photometric data
were obtained with the UNSW Automated Patrol Telescope (APT) and with the 1m
telescope; for spectroscopy, we used the 2.3m telescope, equipped with the
Double Beam Spectrograph. All radial velocities were determined via
cross-correlation, using IAU radial velocity standards.

\section{Results}

\subsection{CY Aqr} 

5 nights of photometry in 2004 and 2 nights of spectroscopy in 2003 and 2004
revealed: (1) the lack of a secondary period, indicating a light curve that has
remained stable over many decades. (2) a stable radial velocity curve, which
agrees very well with that of Fernley et al. (1987), but has better accuracy.
The absence of a change in mean velocity is consistent with the predicted
low-mass companion from the O--C diagram (Fu \& Sterken 2003).

\subsection{ZZ Mic}

We obtained the first RV curve of the star with simultaneous BV light curves.
Three nights of photometry confirmed the double-mode nature of ZZ Mic:
P$_{0}$=0.0671 d and P$_{1}$=0.0522 d. We note, however, that the second mode has
much lower amplitude (A$_{1}$=0.015 mag compared to A$_{0}$=0.189 mag). The
$\gamma$-velocity of ZZ Mic was determined as $-7.8\pm1$ km/s.

\subsection{AD CMi}

We measured an accurate RV curve to search for changes in the systematic
velocity. Our $\gamma$-velocity (35km/s) is in good agreement with  Abhyankar
(1959) (34.5 km/s), which is the only other measurement  published. Our result
does not contradict binarity because the O--C analysis suggested a low-mass
companion that would cause only tiny velocity shift.

\subsection{RS Gru}

We have been monitoring RS Gru since 2003 October. Two sample RV curves are
shown in Fig. 1, where the shift in the systemic velocity is evident. Our
data do not have enough coverage of the orbital phase to determine orbital
period, but we continue collecting further RV measurements.

\begin{figure}[t!]
\resizebox{60mm}{!}{\includegraphics[clip=true]{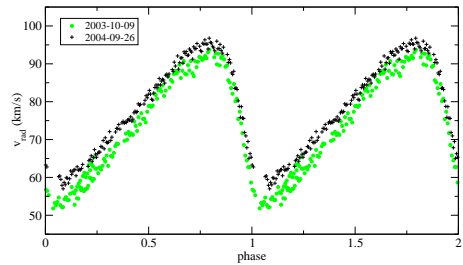}}
\caption{\footnotesize RV curve of RS Gru phased with P=0.147d.}
\label{eta}
\end{figure}

\subsection{BQ Ind}

We observed the star on six consecutive nights in 2004, confirming the presence
of double-mode pulsations. Our data yielded the same frequencies 
($f_{1}$=12.192 c/d, $f_{2}$=15.768 c/d) as those in Sterken et al. (2003),
with no further frequencies in the residuals.

\subsection{RY Lep}

CCD images were taken on 20 nights in 2004. We identify two main  frequencies
as f$_{1}$=4.4416 c/d and f$_{2}$=6.5987 c/d (same as identified in Rodr\'iguez
et al. 2004) and their various linear combinations (up to 10 frequencies).

We also obtained radial velocity measurements on four nights in 2004. The data
revealed that there was a $\sim$25 km/s $\gamma$-velocity shift over the 8
months of observations. This is the first hard evidence for binarity in RY Lep,
and the large $\gamma$-velocity change suggests a relatively massive companion
(comparable to RY Lep).

\begin{acknowledgements}
\begin{scriptsize}
This work has been supported by the Australian Research Council, the
International Postgraduate Research Scholarship (IPRS) programme of the 
Australian Department of Education, Science and Training and Hungarian OTKA
Grant \#T042509. L.L. Kiss is supported by a University of Sydney Postdoctoral
Research Fellowship. The NASA ADS Abstract Service was used to access data and
references.
\end{scriptsize}
\end{acknowledgements}

\bibliographystyle{aa}

\end{document}